\documentclass[
reprint,
amsmath,amssymb,
aps,
prl,
]{revtex4-1}

\usepackage{graphicx}
\usepackage{array}
\usepackage{dcolumn}
\usepackage{bm}
\usepackage{listings}
\usepackage{scalefnt}
\usepackage{xcolor}
\usepackage{booktabs}
\usepackage{multirow}
\usepackage{tensor}
\usepackage{hyperref}
\usepackage{mathtools}
\usepackage[a]{esvect}
\usepackage{textcomp}
\hypersetup{colorlinks=true, linkcolor={blue!80!black}, citecolor={blue!80!black}, urlcolor={blue!80!black}}
\usepackage{color}

\usepackage{physics}
\usepackage[pagewise]{lineno}
\usepackage{mathrsfs}
\usepackage{etoolbox}

\begin{document}

\title{Simultaneous Trapping of Two Optical Pulses in an Atomic Ensemble as Stationary Light Pulses}

\author{U-Shin Kim}
\email{usinkimphysics@gmail.com}

\author{Yoon-Ho Kim}
\email{yoonho72@gmail.com}

\affiliation{Department of Physics, Pohang University of Science and Technology (POSTECH), Pohang 37673, Korea}

\date{\today}

\begin{abstract}
The stationary light pulse (SLP) refers to a zero-group-velocity optical pulse in an atomic ensemble prepared by two counter-propagating driving fields. Despite the uniqueness of an optical pulse trapped within an atomic medium without a cavity, observations of SLP so far have been limited to trapping a single optical pulse due to the stringent SLP phase-matching condition, and this has severely hindered the development of SLP-based applications. In this paper, we first show theoretically that the SLP process  in fact supports two phase-matching conditions and we then utilize the result to experimentally demonstrate simultaneous SLP trapping of two optical pulses for the duration from 0.8 $\mu$s to 2.0 $\mu$s. The characteristic dissipation time, obtained by the release efficiency measurement from the SLP trapping state, is 1.22 $\mu$s, which corresponds to an effective Q-factor of $2.9\times10^{9}$.  Our work is expected to bring forth interesting SLP-based applications, such as, efficient photon-photon interaction, spatially multi-mode coherent quantum memory, creation of exotic photonic gas states, etc. 
\end{abstract}

\maketitle


A stationary light pulse (SLP) is a zero-group-velocity optical pulse trapped in an atomic ensemble prepared by two counter-propagating classical driving fields \cite{Hansen07,Otterbach09, Nikoghosyan09}. In contrast to the light storage based on electromagnetically induced transparency (EIT) in which the photonic excitation is fully mapped to the atomic coherence \cite{Fleishhauer02,Liu01,Phillips01, Cho10a,Cho10b,Cho16},  the SLP process effectively traps the optical pulse in the atomic ensemble while retaining its electromagnetic field nature \cite{Hansen07,Otterbach09, Nikoghosyan09,Zimmer08, Yan12,Iakoupov16}. In the experiment, SLP for a classical light pulse has been reported in a warm atomic vapor system \cite{Bajcsy03} and in a cold atomic ensemble \cite{Lin09,Everett17}. Recently, based on a cold atomic ensemble, quantum SLP has been demonstrated \cite{Park18} and the formation of quantum SLP from a free-propagating single-photon has also been reported \cite{Kim22}. 

The fact that a zero-group-velocity optical pulse can be trapped in an atomic ensemble without the need for a cavity has brought forth to proposals on novel SLP-based applications, such as,  efficient cavity-free nonlinear optics \cite{Andre05, Chen12, Murray17}, photonic quantum gates \cite{Iakoupov18}, fermionization of polaritons \cite{Chang08}, mimicking interacting relativistic theories with SLPs \cite{Angelakis13}, Bose-Einstein condensation of stationary-light polaritons \cite{Fleischhauer08}, etc. Although the majority of such proposed SLP-based applications require interactions of multiple SLPs occupying different spatial modes, observations of SLP so far have been limited to trapping a single optical pulse due to the stringent SLP phase-matching condition \cite{Zhou11}, which has been one of the bottlenecks for realizing various theoretical proposals on SLP-based physics. Extending the SLP physics to support multiple spatial modes would therefore promote the development of, in addition to the above-mentioned potential applications, multiplexed quantum memory, optical image buffer based on SLP, and efficient photon-photon nonlinear optics.

In this work, we demonstrate the simultaneous SLP trapping of two optical pulses in an atomic ensemble.  We first show in theory that the SLP process in fact supports two phase-matching conditions and we then utilize the result to experimentally demonstrate simultaneous SLP trapping of two optical pulses for the duration from 0.8 $\mu$s to 2.0 $\mu$s.  The formation of SLPs is verified by comparing the probe waveforms of SLP with those of EIT. We have also carried out the release efficiency measurement from the SLP trapping state, and the resulting characteristic dissipation time 1.22 $\mu$s, which corresponds to an effective Q-factor of $2.9\times10^{9}$.  

Propagation of the photon-atom excitation through the atomic ensemble is described by a quasi-particle $\hat{\psi}$ described as  \cite{Moiseev14},
\begin{equation}
\hat{\psi}=[ \mathcal{\hat{E}}^{+}\mathrm{cos}\phi+ \mathcal{\hat{E}}^{-}\mathrm{sin}\phi]\mathrm{cos}\theta-\hat{\mathcal{S}}\mathrm{sin}\theta,
\end{equation}
where $\mathcal{\hat{E}}^{+}$ and $\mathcal{\hat{E}}^{-}$ are the slowly varying field operators for the forward-propagating and backward-propagating optical fields, respectively, and $\hat{\mathcal{S}}$ is the collective atomic spin operator. Here, $\mathrm{tan}\phi=\Omega_{\mathrm{BWC}}/\Omega_{\mathrm{FWC}}$, where $\Omega_{\mathrm{FWC}}$ and $\Omega_{\mathrm{BWC}}$ respectively are the Rabi frequencies due to the forward coupling (FWC) and backward coupling (BWC) classical driving fields for the atomic medium, and $\mathrm{tan}^{2}\theta=g^{2} n / \Omega^{2}$, where  $g$ is the atom-photon coupling constant, $n$ is the number of atoms, and $\Omega^{2}=\Omega_{\mathrm{FWC}}^{2}+\Omega_{\mathrm{BWC}}^{2}$. The polariton propagates through the atomic medium with the group velocity $v_g=c_{0} \mathrm{cos}^{2}\theta  \mathrm{cos}2\phi$, where $c_{0}$ is the speed of light in vacuum \cite{Otterbach14}. Note that, when the  Rabi frequencies due to FWC and BWC driving fields are balanced, i.e., $\phi=45^{\circ}$, the SLP state with $v_g=0$ is formed within the atomic medium.

\begin{figure}[t]
\centering
\includegraphics[width=3.2in]{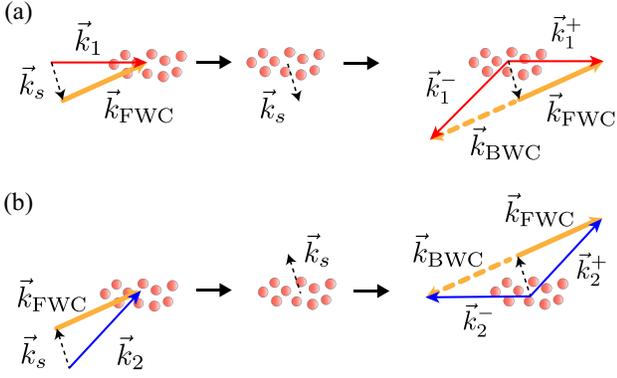}
\caption{(a) Typical phase-matching condition of the SLP trapping process. The momentum conservation conditions, $\vec{k}^{+}_{1} = \vec{k}_{s} +\vec{k}_{\mathrm{FWC}}$ and $\vec{k}^{-}_{1} = \vec{k}_{s} + \vec{k}_{\mathrm{BWC}}$, and energy conservation condition, $|\vec{k}^{+}_{1}| = |\vec{k}^{-}_{1}|$, must be simultaneously satisfied, leading to the angle restriction between $\vec k_1$  and $\vec{k}_{\mathrm{FWC}}$. 
(b) The mirror symmetry with respect to $\vec{k}_{\mathrm{FWC}}$ helps to identify another equally-likely SLP phase-matching condition, enabling simultaneous trapping of two optical pulses. Note, however, that the three vectors $\vec k_1$, $\vec k_2$, and $\vec{k}_{\mathrm{FWC}}$ need not lie on the same plane.} 

\label{fig1}
\end{figure}

In addition to balancing FWC and BWC Rabi frequencies, the formation of SLP from an optical pulse injected into the atomic medium requires a stringent phase-matching condition, see Fig.~\ref{fig1}(a). First, a probe pulse, $\vec{k}_{1}$,  injected into the atomic medium dressed by FWC, $\vec{k}_{\mathrm{FWC}}$, is fully mapped into the collective atomic spin state $\hat{\mathcal{S}}$ with the wave vector $\vec{k}_{s}=\vec{k}_{1}-\vec{k}_{\mathrm{FWC}}$ by adiabatically turning off FWC, which corresponds to the EIT storage process \cite{Fleischhauer00}. If FWC is turned back on after a specified storage time, the propagating optical pulse is retrieved from the collective atomic spin state. For the SLP trapping of the retrieved pulse,  BWC is simultaneously turned on with FWC.  At this step, the wave vectors of trapped forward-propagating field, $\vec{k}^{+}_{1}$, and backward-propagating field, $\vec{k}^{-}_{1}$, must simultaneously satisfy the following conditions: $\vec{k}^{+}_{1} = \vec{k}_{s} +\vec{k}_{\mathrm{FWC}}$ and $\vec{k}^{-}_{1} = \vec{k}_{s} + \vec{k}_{\mathrm{BWC}}$ for the momentum conservation, and $|\vec{k}^{+}_{1}| = |\vec{k}^{-}_{1}|$ for the energy conservation. This phase-matching condition determines the injection angle between probe 1 and FWC to be 0.345$^{\circ}$ for our experimental conditions \cite{Park18,Kim22}, in stark contrast to EIT storage which does not require any phase-matching for the incoming probe beams \cite{Cho10b}.  

Simultaneous trapping of two SLPs requires another input angle that satisfies the phase-matching condition and it can be found by considering the mirror symmetry with respect to  $\vec{k}_{\mathrm{FWC}}$. As shown in Fig.~\ref{fig1}(b), a second probe pulse $\vec{k}_{2}$ injected into the atomic medium symmetrically with respect to   $\vec{k}_{\mathrm{FWC}}$ can create the phase-matching condition which is a mirrored image of the typical SLP phase-matching condition shown in Fig.~\ref{fig1}(a). Note that the wave vector for the spin-wave in the case of Fig.~\ref{fig1}(b) is opposite to the case of Fig.~\ref{fig1}(a). Therefore, for simultaneous SLP trapping of two optical pulses,  the second probe pulse, $\vec{k}_{2}$, must be injected into the atomic medium at the angle of $-0.345^{\circ}$ with respect to  $\vec{k}_{\mathrm{FWC}}$, while the first probe pulse, $\vec{k}_{1}$, is injected at the angle of   $+0.345^{\circ}$ with respect to  $\vec{k}_{\mathrm{FWC}}$.

We note that the phase-matching condition shown in Fig.~\ref{fig1} can support the formation of multi-mode SLPs for more than two optical pulses because the conditions Fig.~\ref{fig1}(a) and Fig.~\ref{fig1}(b) can be satisfied separately. As the only restriction is the angles between the probe pulses $\vec{k}_{1}$ and $\vec{k}_{2}$ and the FWC beam $\vec{k}_{\mathrm{FWC}}$,  the wave vectors $\vec{k}_{1}$ and $\vec{k}_{2}$ form a cone around $\vec{k}_{\mathrm{FWC}}$. The three vectors need not line on the same plane. This feature can be utilized to enable multi-mode photon-photon interaction for interesting quantum information applications, somewhat analogous to multi-mode EIT quantum memory \cite{pu17}. In our experiment, we demonstrate the pair-wise formation of SLPs on the plane  parallel to the optical table.

\begin{figure}[t]
\centering
\includegraphics[width=3.0in]{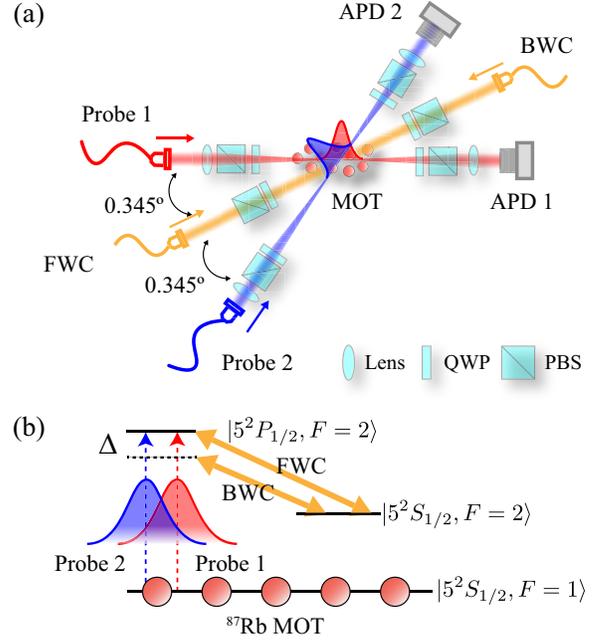}
\caption{(a) Experimental schematic for the simultaneous SLP trapping of two optical pulses. (b) The level diagram of $^{87}$Rb used in the experiment. The detuning $\Delta$=$2\pi\times4$ MHz is introduced to suppress higher-order Raman excitations, which hinders the formation of SLP. APD: avalanched photo-diode, FWC: forward coupling beam, BWC: backward coupling beam, MOT: magneto-optical trap.}
\label{fig2}
\end{figure}

The schematic of the experiment to demonstrate simultaneous SLP trapping of two optical pulses is shown in Fig.~\ref{fig2}(a). We exploit a cigar-shaped $^{87}$Rb cold atomic ensemble consisting of a pair of rectangular anti-Helmholtz coils and 1-inch-diameter trapping beams of 130 mW \cite{Cho14,Lee16,Park17,Zhao19}. The duration of each experimental sequence is 200 ms, of which 190 ms is used for preparing the atomic medium starting with the plain magneto-optical trap (MOT) loading phase of 173.5 ms. It is then followed by the temporal dark MOT phase of 15 ms to trap more atoms by gradually ramping down the trapping beam powers. The decreased beam powers help to increase the optical depth (OD) of the atomic ensemble by preventing the trapped atoms from reheating by the intense trapping beams \cite{Ketterle92, Hsiao14}. At the start of the temporal dark MOT phase, the magnetic field compression phase is simultaneously initiated by ramping up the current of the anti-Helmholtz coils from 5.8 A to 13.6 A (SM100-AR-75, DELTA ELEKTRONIKA). After 13 ms of the magnetic field compression phase, the anti-Helmholtz coils are then switched off using an insulated-gate bipolar transistor (IGBT) (1MBI400V-120-50, Fuji Electronic) to remove residual magnetic field stray. Finally, the atoms are prepared in the ground state $|5^{2}S_{1/2}, F=1\rangle$ by shining a 1.5 ms optical pulse which is resonant to $|5^{2}S_{1/2}, F=2\rangle$ to $|5^{2}P_{3/2}, F=2\rangle$ transition. The SLP trapping experiment is carried out during the experimental window of 10 ms and the relevant energy level diagram is shown in Fig.~\ref{fig2}(b).  The probes and the coupling beams are prepared from a 795 nm-centered external-cavity diode laser (ECDL) combined with a tapered amplifier system (TA Pro, Toptica). The ECDL is locked at the 120 MHZ red-detuned frequency from $|5^{2}S_{1/2}, F=2\rangle$ to $|5^{2}P_{1/2}, F=2\rangle$ transition. FWC and BWC are then modulated using acousto-optic modulators (AOM) (ATM-1201A2, IntraAction) for the experimental sequence by 120 MHz and 116 MHz, respectively. Probe 1 and probe 2 are prepared by up-shifting the optical frequency of the ECDL using an electro-optic modulator (EOM) (Visible Phase Modulator 4851, New Focus) by 6.79 GHz. Finally, the probes are modulated using AOMs (ATM-801A2, IntraAction) by 80 MHz to have a 2 $\mu$s temporal width and the frequency resonant to $|5^{2}S_{1/2}, F=1\rangle$ to $|5^{2}P_{1/2}, F=2\rangle$ transition as shown in Fig.~\ref{fig2}(b). As a result, the atomic ensemble prepared for our experiment has an OD of 60, the dephasing rate between the two ground states of $2\pi\times 60$ kHz, and the phase mismatch for SLP process, $\Delta_k L$, of 0.05.

For the available 10 ms experimental window, for optimal results, we use the first 10 $\mu$s for the SLP experiment. First, probe 1 and probe 2 with 2 $\mu$s temporal widths are simultaneously injected to the atomic ensemble dressed by FWC with the phase-matching angle $\pm0.345^{\circ}$ with respect to $\vec{k}_{\mathrm{FWC}}$ as shown in Fig.~\ref{fig2}(a). Probe 1 and probe 2 are then mapped into the collective spin states by adiabatically turning off FWC, which corresponds to the EIT storage process. After the 2 $\mu$s EIT storage process, FWC and BWC are turned on simultaneously to form SLPs of probe 1 and probe 2 inside the atomic ensemble. The detuning between FWC and BWC, $\Delta=2\pi\times4$ MHz, is introduced to avoid higher-order Raman excitations, which hinder the formation of SLP \cite{Wu10}. After a specified SLP trapping duration, the probe pulses are released from the SLP states by turning off BWC and detected by two avalanched photo-diodes (APDs) (APD120A/M, Thorlabs). All traces from the APDs are acquired by averaging 50 experimental sequences using an 1 GHz-bandwidth oscilloscope (DPO7104, Tektronix).

\begin{figure}[t]
\centering
\includegraphics[width=3.1in]{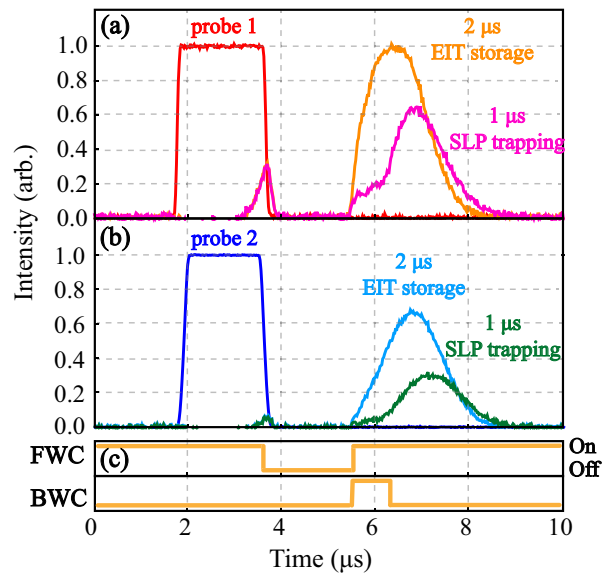}
\caption{EIT storage versus SLP trapping for (a) probe 1 and (b) probe 2. EIT storage occurs by turning off FWC for 2 $\mu$s. When only FWC is turned back on, probe pulses are immediately retrieved, but if both FWC and BWC are turned on simultaneously, photon retrieval is suppressed and delayed until BWC is turned off, signaling SLP formation. 
(c) Temporal sequence of FWC and BWC laser beams.
}
\label{fig3}
\end{figure}

Figure~\ref{fig3} shows, for probe 1 and probe 2,  the experimental results of the EIT storage of  2 $\mu$s only and those of the EIT storage of  2 $\mu$s followed by the SLP trapping of 1 $\mu$s. The prepared cold atom ensemble (for the duration of 190 ms prior to the experimental window)  is dressed with FWC and, at the time of 1.6 $\mu$s,  probe 1 and probe 2 are injected into the atomic medium as shown in Fig.~\ref{fig2}. Due to EIT formed by FWC, when FWC is turned off between 3.6 $\mu$s and 5.6 $\mu$s, probe 1 and probe 2 are simultaneously mapped to the collective atomic spin states.  After the EIT storage of 2 $\mu$s, FWC is turned back on at the time of 5.6 $\mu$s to retrieve the two probe pulses as propagating optical pulses from the collective atomic spin states, as evidenced in the EIT traces in Fig.~\ref{fig3}. If both FWC and BWC are turned on simultaneously at 5.6 $\mu$s to 6.6 $\mu$s, the retrieved optical pulses are immediately trapped within the atomic medium as SLPs, and, when BWC is later turned off, SLPs are converted into propagating optical pulses. The suppressed emission during the time both FWC and BWC are simultaneously turned on and the subsequent emission of the optical pulses when BWC is turned off clearly indicate the formation of simultaneous SLPs for probe 1 and probe 2 for the duration of 1 $\mu$s.

The EIT/SLP efficiency differences observed in Fig.~\ref{fig3}(a) and Fig.~\ref{fig3}(b) for probe 1 and probe 2 come from the fact that, in our experimental setup, each probe beam experiences a slightly different OD  and has a different overlap with FWC. This can be confirmed by comparing their efficiencies and group delays during the EIT storage processes shown in Fig.~3. Let us first consider the retrieval efficiencies of the two probe beams after the 2 $\mu$s EIT storage process. The retrieval efficiency of probe 1 is measured to be 80.8\%, and that of probe 2 is measured to be 59.1\%. Since higher OD gives higher retrieval efficiency under the same storage duration, we conclude that probe 1 experiences higher OD than probe 2 does. Let us now compare the group delays (due to the slow-light effect) experienced by the two probe beams after the 2 $\mu$s EIT storage process. The group delay is calculated by subtracting the storage time of 2 $\mu$s from the total time delay measured at the peak of the observed probe waveform. The group delay of probe 1 is measured to be 1.9 $\mu$s, and that of probe 2 is measured to be 2.4 $\mu$s from Fig. 3. In general, higher OD and lower $\Omega_{\mathrm{FWC}}$ result in a larger group delay. Since probe 1 experiences higher OD than probe 2, we can conclude that the spatial overlap between probe 2 and FWC is slightly worse than the spatial overlap between probe 1 and FWC.

\begin{figure}[t]
\centering
\includegraphics[width=3.1in]{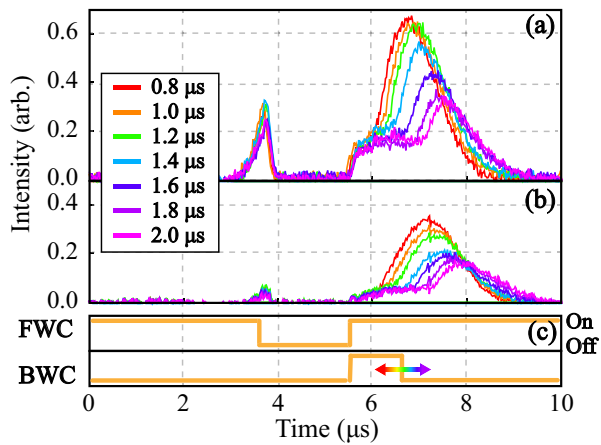}
\caption{Simultaneous SLP trapping of (a) probe 1 and (b) probe 2 from 0.8 $\mu$s to 2.0 $\mu$s. (c) Temporal sequence of FWC and BWC laser beams.}
\label{fig4}
\end{figure}

In Fig.~\ref{fig4}, we show the experimental results of simultaneous SLP trapping of probe 1 and probe 2 for different SLP trapping times by varying the duration of the BWC turn-on time. As before, SLPs are formed for probe 1 and probe 2 after the EIT storage process of 2 $\mu$s. The SLP trapping time is varied from 0.8 $\mu$s to 2.0 $\mu$s in 0.2 $\mu$s steps. The experimental results in Fig.~\ref{fig4} clearly display the suppressed emission of the probes while FWC and BWC are both turned on, and the delayed release of the probe pulses when BWC is turned off after a specified SLP trapping time.

\begin{figure}[t]
\centering
\includegraphics[width=3.1in]{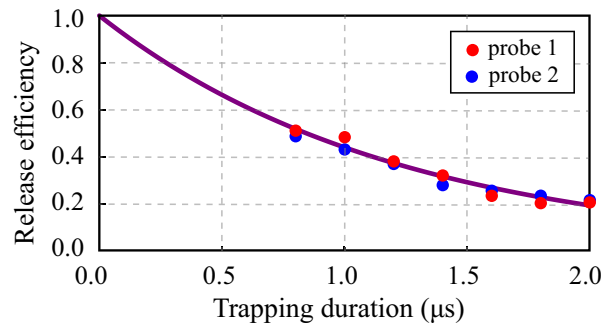}
\caption{The observed SLP trapping efficiency as the trapping times are varied. Both probe 1 and probe 2, as they undergo simultaneous SLP processes, exhibit the same exponential decay with the characteristic dissipation time of 1.22 $\mu$s represented as the solid line. The characteristic dissipation time of 1.22 $\mu$s corresponds to the effective Q-factor of $2.9 \times 10^9$.}
\label{fig5}
\end{figure}

As the SLP trapping times become longer, the peaks of the released optical pulses become smaller as shown in Fig.~\ref{fig4}.  In Fig.~\ref{fig5}, we summarize the release efficiency measurements of probe 1 and probe 2, as a function of the SLP trapping time, from the simultaneous SLP trapping shown in Fig.~\ref{fig4}. The release efficiencies of probe 1 and probe 2 exhibit the same exponential energy dissipation with the characteristic dissipation time of $\tau=1.22$ $\mu$s. Since probe 1 and probe 2 are trapped in the atomic ensemble while retaining their electromagnetic field nature, the simultaneous SLP trapping process can be understood with the analogy of an optical cavity system that confines the electromagnetic field within a restricted volume. The Q-factor of an optical cavity system is typically given by $Q={2\pi f_{0} E}/{P} = 2\pi f_0 \tau$, where $f_{0}$ is the resonant frequency, $E$ is the stored energy in the cavity, $P=-{dE}/{dt}$ is the energy dissipation rate, and $\tau$ is the characteristic energy dissipation time \cite{Saleh07,anderson84}. The characteristic  dissipation time of $\tau=1.22$ $\mu$s corresponds to the optical cavity system with an effective Q-factor of $2.9\times10^{9}$ \cite{Chen12}. Note that the effective Q-factor of $2.9\times 10^{9}$ is comparable to other ultrahigh Q-factor optical cavity systems such as the silicon or crystalline resonator system.

The high effective Q-factor of the SLP trapping process demonstrated in this work implies that there is a strong atom-photon coupling.  The atom-photon coupling can be estimated by $N$-atom cooperativity, $C_{N}$, as the photons are effectively trapped within the atomic ensemble, similarly to a cavity quantum electrodynamics (QED) system. The $N$-atom cooperativity is given as $C_{N}=4g^{2}N/\kappa\Gamma$, where $g$ is the atom-field coupling constant, $\kappa$ is the effective linewidth of the cavity, $\Gamma$ is the decay rate of the atom, and $N$ is the number of atoms \cite{beck16, beck14}. In our SLP experimental setup, the relevant parameters are estimated as $\{g, \kappa, \Gamma\}=2\pi\times\{0.24, 0.13, 5.8\}\ \mathrm{MHz}$ from the experimental parameters (the optical depth $\mathrm{OD}=60$, the group delay $\tau_{g}\approx 2\ \mu\mathrm{s}$, the Q-factor $Q=2.9\times10^{9}$, and the length of atomic ensemble $L=10\ \mathrm{mm}$) using the relations $\tau_{g}=\mathrm{OD}\times \Gamma / |\Omega_{\mathrm{FWC}}|^2$ \cite{fleischhauer05} and $v_{g}={L}/{\tau_{g}}={c_{0} |\Omega_{\mathrm{FWC}}|^2}/({|\Omega_{\mathrm{FWC}}|^2+g^{2}N})$ \cite{Fleischhauer00}.  This leads to the $N$-atom cooperativity of our SLP system as $C_{N}\sim8\times10^{6}$, which is larger by an order of magnitude than those reported in recent cavity QED experiments \cite{beck16, beck14, chen13, hosseini16}.

In summary, we have for the first time demonstrated simultaneous SLP trapping of two optical pulses in a cold atomic ensemble. We have shown theoretically that the SLP process in fact supports two phase-matching geometries, and experimentally demonstrated simultaneous SLP trapping of two optical pulses for the duration from 0.8 $\mu$s to 2.0 $\mu$s. Although the SLP trapping processes for the two optical pulses showed slightly different efficiencies due to the experimental geometry, both SLP trapping processes exhibited the same characteristic dissipation time of 1.22 $\mu$s, which corresponds to an effective Q-factor of $2.9\times 10^{9}$. As our SLP system reports a very large effective  $N$-atom cooperativity, $C_{N}\sim8\times10^{6}$, our work is expected to bring forth interesting SLP-based applications, such as, cavity-free nonlinear optics \cite{Andre05, Chen12, Murray17}, photonic quantum gates \cite{Iakoupov18}, fermionization of polaritons \cite{Chang08}, mimicking relativistic theories with SLPs \cite{Angelakis13}, and Bose-Einstein condensates of stationary-light polaritons \cite{Fleischhauer08}. 


This work was supported in part by the National Research Foundation of Korea (2019R1A2C3004812) and the ITRC support program (IITP-2022-2020-0-01606).



\begin{thebibliography}{}

\bibitem{Hansen07} K. R. Hansen and K. M\o lmer, Phys. Rev. A \textbf{75}, 053802 (2007).

\bibitem{Otterbach09} J. Otterbach, R. G. Unanyan, and M. Fleischhauer, Phys. Rev. Lett. \textbf{102}, 063602 (2009).

\bibitem{Nikoghosyan09} G. Nikoghosyan and M. Fleischhauer, Phys. Rev. A \textbf{80}, 013818 (2009).

\bibitem{Fleishhauer02} M. Fleischhauer and M. D. Lukin, Phys. Rev. A \textbf{65}, 022314 (2002).

\bibitem{Liu01} C. Liu, Z. Dutton, C. H. Behroozi, and L. V. Hau,  Nature \textbf{409}, 490-493 (2001).

\bibitem{Phillips01} D. F. Phillips, A. Fleischhauer, A. Mair, R. L. Walsworth, and M. D. Lukin, Phys. Rev. Lett. \textbf{86}, 783 (2001).

\bibitem{Cho10a} Y.-W. Cho and Y.-H. Kim, Opt. Express \textbf{18}, 25786 (2010).

\bibitem{Cho10b} Y.-W. Cho and Y.-H. Kim, \pra \textbf{82}, 033830 (2010).

\bibitem{Cho16} Y.-W. Cho, G. T. Campbell, J. L. Everett, J. Bernu, D. B. Higginbottom, M. T. Cao, J. Geng, N. P. Robins, P. K. Lam, and B. C. Buchler, Optica \textbf{3}, 100-107 (2016).

\bibitem{Zimmer08} F. E. Zimmer, J. Otterbach, R. G. Unanyan, B. W. Shore, and M. Fleischhauer, Phys. Rev. A \textbf{77}, 063823 (2008).

\bibitem{Yan12} Y. Zhang, Y. Zhang, X.-H. Zhang, M. Yu, C.-L. Cui, J.-H. Wu, Phys. Lett. A \textbf{376}, 656-661 (2012).

\bibitem{Iakoupov16} I. Iakoupov, J. R. Ott, D. E. Chang, and A. S. S\o rensen, Phys. Rev. A \textbf{94}, 053824 (2016).

\bibitem{Bajcsy03} M. Bajcsy, A. S. Zibrov, and  M. D. Lukin, Nature \textbf{426}, 638-641 (2003).

\bibitem{Lin09} Y.-W. Lin, W.-T. Liao, T. Peters, H.-C. Chou, J.-S. Wang, H.-W. Cho, P.-C. Kuan, and I. A. Yu, Phys. Rev. Lett. \textbf{102}, 213601 (2009).

\bibitem{Everett17} J. L. Everett, G. T. Campbell, Y.-W. Cho, P. V.-Gris, D. B. Higginbottom, O. Pinel, N. P. Robins, P. K. Lam and B. C. Buchler, Nat. Phys. \textbf{13}, 68-73 (2017).

\bibitem{Park18} K.-K. Park, Y.-W. Cho, Y.-T. Chough, and Y.-H. Kim, Phys. Rev. X \textbf{8}, 021016 (2018).

\bibitem{Kim22} U.-S. Kim, Y. S. Ihn, C.-H. Lee, and Y.-H. Kim, AVS Quantum Sci. \textbf{4}, 021403 (2022).

\bibitem{Andre05} A. Andr\'{e}, M. Bajcsy, A. S. Zibrov, and M. D. Lukin, Phys. Rev. Lett. \textbf{94}, 063902 (2005).

\bibitem{Chen12} Y.-H. Chen, M.-J. Lee, W. Hung, Y.-C. Chen, Y.-F. Chen, and I. A. Yu, Phys. Rev. Lett. \textbf{108}, 173603 (2012).

\bibitem{Murray17} C. R. Murray and T. Pohl, Phys. Rev. X \textbf{7}, 031007 (2017).

\bibitem{Iakoupov18} I. Iakoupov, J. Borregaard, and A. S. S\o renson, Phys. Rev. Lett. \textbf{120}, 010502 (2018).

\bibitem{Chang08} D. E. Chang , V. Gritsev, G. Morigi, V. Vuleti\'{c}, M. D. Lukin, and E. A. Demler, Nat. Phys.  \textbf{4}, 884-889 (2008).

\bibitem{Angelakis13} D. G. Angelakis, M.-X. Huo, D. Chang, L. C. Kwek, and V. Korepin, Phys. Rev. Lett.  \textbf{110}, 100502 (2013).

\bibitem{Fleischhauer08} M. Fleischhauer, J. Otterbach, and R. G. Unanyan, Phys. Rev. Lett.  \textbf{101}, 163601 (2008).

\bibitem{Zhou11} H.-T. Zhou, D.-W. Wang, D. Wang, J.-X. Zhang, and S.-Y. Zhu, Phys. Rev. A  \textbf{84}, 053835 (2011).

\bibitem{Moiseev14} S. A. Moiseev, A. I. Sidorova, and B. S. Ham, Phys. Rev. A \textbf{89}, 043802 (2014).

\bibitem{Otterbach14} J. Otterbach, J. Ruseckas, R.G. Unanyan, G. Juzeli$\mathrm{\bar{u}}$nas, and M. Fleischhauer, Phys. Rev. Lett. \textbf{104}, 033903 (2010).

\bibitem{Fleischhauer00} M. Fleischhauer and M. D. Lukin, Phys. Rev. Lett. \textbf{84}, 5094 (2000).

\bibitem{pu17} Y.-F. Pu, N. Jiang, W. Chang, C. Li, and L.-M. Duan, Nat. Commun. \textbf{8}, 15359 (2017). 

\bibitem{Cho14} Y.-W. Cho, K.-K. Park, J.-C. Lee, and Y.-H. Kim, Phys. Rev. Lett. \textbf{113}, 063602 (2014).

\bibitem{Lee16} J.-C. Lee, K.-K. Park, T.-M. Zhao, and Y.-H. Kim, Phys. Rev. Lett. \textbf{117}, 250501 (2016).

\bibitem{Park17} K.-K. Park, Y.-W. Cho, J.-H. Kim, T.-M. Zhao, and Y.-H. Kim, Optica \textbf{4}, 1293 (2017).

\bibitem{Zhao19} T.-M. Zhao, Y. S. Ihn, and Y.-H. Kim, Phys. Rev. Lett. \textbf{122}, 123607 (2019).

\bibitem{Ketterle92} W. Ketterle, K. B. Davis, M. A. Joffe, A. Martin, and D. E. Pritchard, Phys. Rev. Lett. \textbf{70}, 2253 (1993)

\bibitem{Hsiao14} Y.-F. Hsiao, H.-S. Chen, P.-J. Tsai, and Y.-C. Chen, Phys. Rev. A \textbf{90}, 055401 (2014).

\bibitem{Wu10} J.-H. Wu, M. Artoni, and G. C. La Rocca, Phys. Rev. A \textbf{82}, 013807 (2010).

\bibitem{Saleh07} B. E. A. Saleh and M. C. Teich, \textit{Fundamentals of photonics} (Wiley-Interscience, New Jersey, 2007), 2nd ed., p. 376. 

\bibitem{anderson84} D. Z. Anderson, J. C. Frisch, and C. S. Masser, Appl. Opt. \textbf{23}, 1238-1245 (1984).

\bibitem{beck16} K. M. Beck, M. Hosseini, Y. Duan, and V. Vuleti\'c, PNAS \textbf{113}, 35 (2016).

\bibitem{beck14} K. M. Beck, W. Chen, Q. Lin, M. Gullans, M. D. Lukin, and V. Vuleti\'c, Phys. Rev. Lett. \textbf{113}, 113603 (2014).

\bibitem{fleischhauer05} M. Fleischhauer, A. Imamoglu, and J. P. Marangos, Rev. Mod. Phys. \textbf{77}, 2 (2005).

\bibitem{chen13} W. Chen, K. M. Beck, R. B\"ucker, M. Gullans, M. D. Lukin, H. T.-Suzuki, and V. Vuleti\'c, Science \textbf{341}, 768-770 (2013).

\bibitem{hosseini16} M. Hosseini, K. M. Beck, Y. Duan, W. Chen, and V. Vuleti\'c, Phys. Rev. Lett. \textbf{116}, 033602 (2016).

 \end{thebibliography}
\end{document}